\documentclass{jetpl}
\twocolumn

\usepackage[english]{babel}
\usepackage[T1]{fontenc}
\usepackage[utf8]{luainputenc}
\usepackage{amsmath}
\usepackage{graphicx}
\usepackage{xcolor}

\lat

\usepackage{bm}
\def\B{\bm{{\rm B}}}
\def\S{\bm{{\rm S}}}
\def\X{\bm{{\rm X}}}
\def\n{\bm{{\rm n}}}
\def\bsigma{\bm{\sigma}}
\def\bomega{\bm{\omega}}
\def\env{{\rm env}}
\def\od{{\uparrow\downarrow}}

\title{Non-adiabatic geometric phases and dephasing in an open quantum system}

\rtitle{Non-adiabatic geometric phases and dephasing in an open quantum system}

\sodtitle{Non-adiabatic geometric phases and dephasing in an open quantum system}

\author{A.E.~Svetogorov$^{+*}$ and
Yu.~Makhlin$^{+*}$\/\thanks{e-mail: svetogorov@itp.ac.ru, makhlin@itp.ac.ru}}

\rauthor{A.E.~Svetogorov, Yu.~Makhlin}

\address{$^+$L.D.Landau Institute for Theoretical Physics, acad. Semyonov 
av. 1a, 142432, Chernogolovka, Russia\\
$^*$ Moscow Institute of Physics and Technology, 141700, Dolgoprudny, Russia}

\abstract{%
We analyze the influence of a dissipative environment on geometric phases in a quantum system
subject to non-adiabatic evolution. We find dissipative contributions to the acquired phase
and modification of dephasing, considering the cases of weak short-correlated noise as well as of
slow quasi-stationary noise. Motivated by recent experiments, we find the leading non-adiabatic
corrections to the results, known for the adiabatic limit.}
\begin{document}
\maketitle

The Berry phase~\cite{Berry} is a celebrated instance of geometric phases in physics~\cite{Shapere},
which occurs during adiabatic evolution of a quantum system. In the analysis of generalizations of
the Berry phase, Aharonov and Anandan found a geometric phase even for non-adiabatic
evolutions~\cite{AA}. When a quantum system is coupled to an environment, phases, acquired by the
system during its evolution, are modified. In particular, it was shown that in a quantum system
subject simultaneously to adiabatic variation of its parameters and to weak short-correlated
external noise, the phase acquires a geometric environment-induced
contribution~\cite{Whitney,SMYalta}. Furthermore, the environment-induced decoherence is modulated
by the parameter variation, which results in a geometric contribution to
dephasing~\cite{Whitney,Whitney07}. Here we analyze the dynamics of an open quantum system during
non-adiabatic evolution. In this case in a closed system the total phase is a combination of the
dynamical phase and the geometric Aharonov-Anandan phase. We find, how this phase is modified by the
environment. In particular, motivated by recent experiments with superconducting qubits, we study
the adiabatic limit and find the leading non-adiabatic corrections.

The Berry phase was measured directly, in the original setting with cyclic variation of the
magnetic field, in NMR systems~\cite{Suter}. Some time ago the degree of control over the quantum
state and the coherence level allowed for direct observation of the Berry phase in
superconducting qubits~\cite{Leek}. In later experiments, the influence of noise on the Berry phase
in this system was studied and the geometric dephasing was analyzed~\cite{Berger2013,Berger}.

We first describe the coherent Aharonov-Anandan phase and then consider the influence of
dissipation. We consider a quantum two-level system and use the spin-1/2 language for its
description. The Hamiltonian of a two-state system can be presented as $H=-\B(t)\bsigma/2$, where
$\B$ can be referred to as the (pseudo)magnetic field. For any variation of $\B(t)$ over a certain
period, $0<t<T$, the unitary evolution operator has two eigenstates, referred to as {\it cyclic
states} as they return to their initial values up to a phase, i.e., the two corresponding opposite
spin vectors return to their initial directions. The relative phase between these states acquired
over the time $T$ defines the angle of rotation in spin space over the cyclic direction. Aharonov
and Anandan~\cite{AA} showed that it consists of two contributions defined below, a dynamical and
geometric phase. Indeed, let $\S(t)$ be a cyclic state. Consider a spin frame with the $z'$ axis
along $\S(t)$. The magnetic field in this frame is $\B'=\B+\bomega$, where $\bomega$ is the
angular velocity of the rotating frame relative to the lab frame. Since the equation of motion
in this frame reads $\dot{\S}=\S\times\B'$, and $\S$ is  stationary, $\B'$ has the same direction,
$\B'\parallel\S$. Thus the total relative phase between $\S$ and $-\S$, picked during the evolution,
is given by $\int dt B' = \int dt B_\parallel + \int dt \omega_\parallel$, where the subindex
$\parallel$ indicates projection onto $\S$. The first contribution to this total phase, also given
by $\int dt \left\langle\psi\right| H \left|\psi\right\rangle$, the time integral of the average
energy, is the dynamical phase (here $\left|\psi(t)\right\rangle$ is the quantum state,
corresponding to $\S(t)$). The rest, $\int dt \omega_\parallel$, similar to the case of the
adiabatic evolution, is given by the solid angle, subtended by $\S(t)$. This geometric contribution
is the Aharonov-Anandan phase.

%%%%%%%%%%%%%%%%%%%%%%%%%%%%%%%%%%%%%%%%%%%%%%%%%%%%%%%%%%%%%%%%%%
\begin{figure}
\centerline{\includegraphics[scale=0.3]{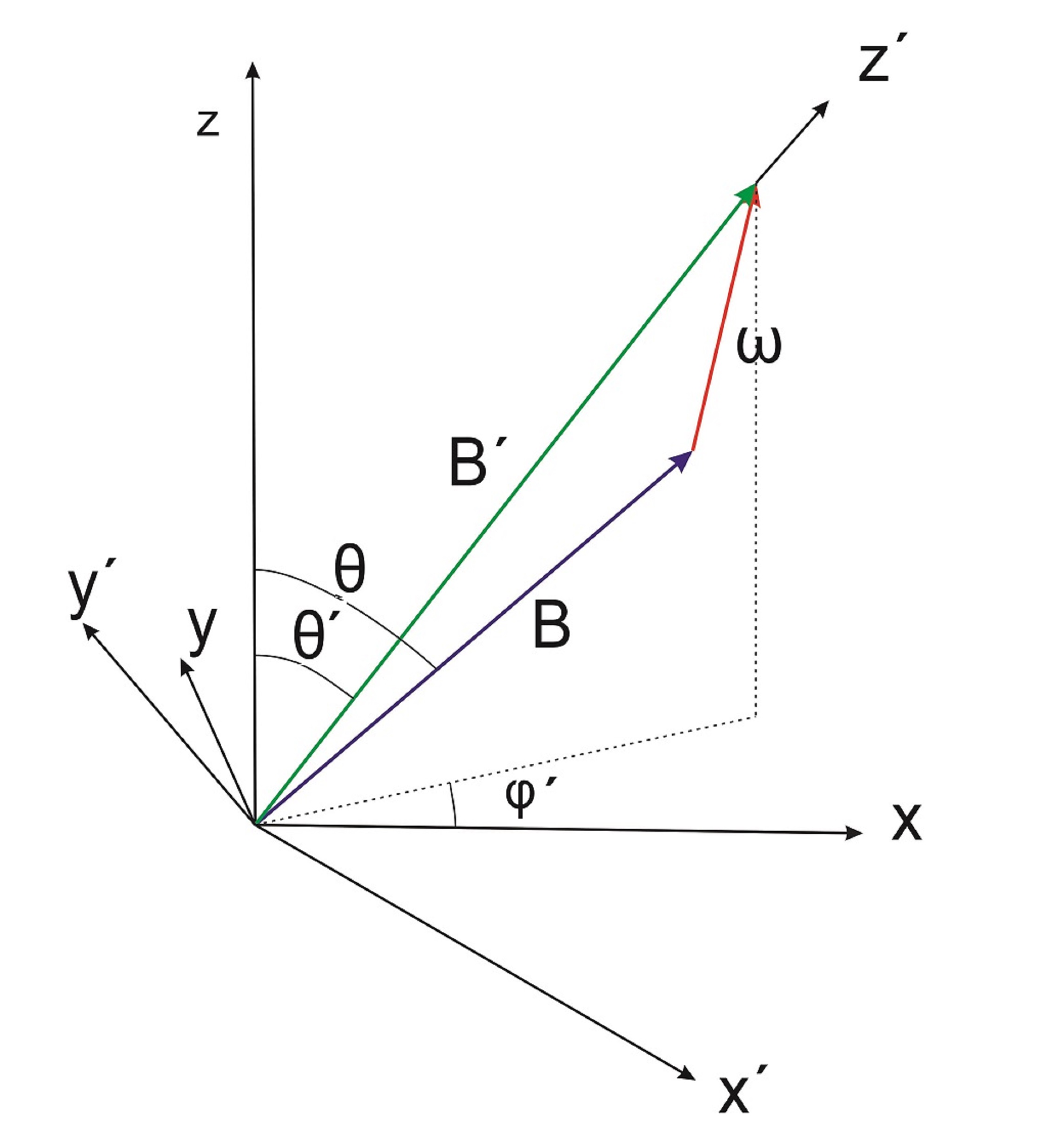}}
\caption{The rotating spin reference frame $x'y'z'$, rotating with angular velocity $\bomega$
relative to the stationary frame $xyz$.}
\end{figure}
%%%%%%%%%%%%%%%%%%%%%%%%%%%%%%%%%%%%%%%%%%%%%%%%%%%%%%%%%%%%%%%%%%

When the system is subject to external noise (coupled to an environment), its dynamics is modified.
For a static field $\B$, dissipation induces relaxation and dephasing, as well as a
modification of the dynamical phase. For weak short-correlated noise, it was shown~\cite{Whitney}
that for an adiabatically slowly varying field, on top of that dissipation modifies the geometric
Berry phase and introduces a geometric contribution to dephasing. Here we extend this analysis to
the case of non-adiabatic evolution. We find the acquired phase and dephasing for a system coupled
to environment and subject to non-adiabatic manipulations.

Consider a quantum system, a spin-1/2 in our case, coupled to an environment:
\begin{equation}
H = -\frac{1}{2}\B\bsigma -\frac{1}{2}\X\bsigma + H_\env,
\end{equation}
where $\X=X\n$ is an operator of the environment, which represents noise experienced by the quantum
system (we assume $\langle X\rangle=0$); $H_\env$ governs the dynamics of the modes of the
environment.
To be specific, we consider unidirectional noise with $\n=\hat z$. Such anisotropy of the
noise is relevant, e.g., for superconducting qubits, where different (pseudo-)spin directions
correspond to different physical variables~\cite{Erice}. We comment on other situations later.

In the rotating frame (RF), with the $z'$-axis chosen along $\bf S$ and the
$y'$-axis
orthogonal to the original $z$-axis (this choice ensures that the frame returns to its initial
state after a cycle), the Hamiltonian reads
\begin{equation}
\hat{H}'=-\frac{1}{2}B'\sigma_{z'} + \hat{H}_{int} + H_{env}
\end{equation}
with the interaction term
\begin{equation*}
\hat{H}_{int}
= -\frac{1}{2}X \left(\cos\theta'\sigma_{z'} -
\sin\theta'\sigma_{x'} \right)
\,,
\end{equation*}
where $\theta'$ is the angle between the direction $z$ of the fluctuations and the $z'$ axis.

The phases can be read off from the off-diagonal element of the density matrix in the RF,
$\langle\sigma_{x'}+i\sigma_{y'}\rangle=\rho_\od$.
Using the real-time Keldysh technique~\cite{Shoeller,Erice},
we can derive the kinetic equation for the density matrix.
For weak noise with a short correlation time $\tau_c\ll T_1,T_2$
(here $T_1, T_2$ are the longitudinal and transverse relaxation times) one can use the
Bloch-Redfield and the rotating-wave approximations to find a closed equation for the off-diagonal
entry $\rho_\od$:
\begin{equation}\label{eq:rhodot}
\frac{d}{dt}\rho_\od(t)=iB'\rho_\od(t)-\Gamma_2\rho_\od(t)
\end{equation}
with the complex `dephasing rate' given by
\begin{eqnarray}
\Gamma_2 &=&
\intop_{-\infty}^{t} dt' S(t-t') \biggl[
\cos\theta'(t) \cos\theta'(t') + \nonumber\\
&+& \frac{1}{2} \sin\theta'(t) \sin\theta'(t')
\exp\left(-i\int_{t'}^{t}B'(\tau)d\tau\right)
\biggr]
\,,
\end{eqnarray}
via the noise correlator $S(t-t') = \frac{1}{2} \langle X(t)X(t') + X(t')X(t) \rangle$.

To the leading order in small $\omega\tau_{c}$ and $\dot{B'}{\tau_c}^2$ 
\begin{eqnarray*}
\Gamma_2
&=&-\intop_0^\infty dt S(t) \{\cos\theta' (\cos\theta'+\sin\theta'\omega_{y'}t)\\
&+&\frac{1}{2}\sin\theta' (\sin\theta'-\cos\theta'\omega_{y'}t) \exp(-iB't)\\
&+&\frac{i}{2}\sin^{2}\theta' \dot{B'}\frac{t^{2}}{2} \exp(-iB't)\}
\end{eqnarray*}
In terms of the noise power spectrum, the Fourier transform of $S(\tau)$, we obtain
\begin{eqnarray}
\Gamma_2  = 
-i\int \frac{d\Omega}{2\pi} S(\Omega)
\left( \frac{\cos^2\theta'}{\Omega+i0}
+ \frac{\sin^2\theta'}{2(\Omega-B'+i0)} \right)
\nonumber\\
+ 
\omega_{y'} \int \frac{d\Omega}{2\pi} S(\Omega)
\left(
\frac{\sin\theta'\cos\theta'}{(\Omega+i0)^2}
-\frac{\sin\theta'\cos\theta'}{2(\Omega-B'+i0)^2} \right)
\nonumber\\
- 
\frac{1}{2}\dot{B'} \sin^2\theta' \int \frac{d\Omega}{2\pi} S(\Omega)
\frac{1}{(\Omega-B'+i0)^{3}}
\,.\label{eq:G2}
\end{eqnarray}

The imaginary part of Eq.~(\ref{eq:G2}) gives the acquired phase:
\begin{equation*}
\Delta\Phi=\int (B' - \mathop{\rm Im}\Gamma_2) dt \,.
\end{equation*}
As one can see, the second and third terms in~(\ref{eq:G2}) vanish after integration over a
closed trajectory.

In a recent experiment~\cite{Berger}, the spin underwent uniform evolution around the $z$-axis
(the magnetic field of fixed magnitude varied circularly around the $z$-axis) in a near-adiabatic
limit. While the noise in Ref.~\cite{Berger} was rather quasi-stationary, for comparison we expand
our result to the second order in $\omega$:
\begin{equation}
\Delta\Phi \approx -\mathop{\rm Re}
\int dt \int \frac{d\Omega}{2\pi} S(\Omega)F(\omega,\theta,\Omega) \,,
\end{equation}
where
\begin{eqnarray}\label{eq:F}
F(\omega,\theta,\Omega) &\approx& \frac{\sin^2\theta}{2(\Omega-B+i0)}
 (1+\beta\omega+\gamma\omega^2) \,,\\
\beta&=&\frac{(3B-2\Omega)\cos\theta}{B(\Omega-B+i0)} \,,
\quad \gamma = \frac{3\cos^2\theta-\sin^2\theta}{B^{2}}\nonumber\\
&+& \frac{\cos^2\theta}{(\Omega-B+i0)^2} + \frac{\sin^2\theta}{2B(\Omega-B+i0)} \,.\nonumber
\end{eqnarray}
Eq.~(\ref{eq:F}) includes the $\omega$-independent dynamical part, the geometric part
$\propto\omega$~\cite{Whitney}, and the leading non-adiabatic
correction $\propto\omega^2$, which is non-geometric.

The real part of Eq.~(\ref{eq:G2}) gives the dephasing rate. Dropping the last two terms,
which vanish after integration over a closed path (the dephasing, however, is well-defined for an
open path too, cf.Ref.~\cite{Whitney}), we find
\begin{equation}\label{eq:T2}
\frac{1}{T_{2}}  =  \frac{1}{2}S(0)\cos^{2}\theta'+\frac{1}{4}S\left(B'\right)\sin^{2}\theta'
\end{equation}

The dynamics of the level occupations, the diagonal entries of the density  matrix, is decoupled
from the phase and describes their relaxation. From the Bloch equations in the rotating frame
we find the relaxation rate
\begin{equation}\label{eq:T1}
\frac{1}{T_1} = \frac{\sin^2\theta'}{2} S(B') \,.
\end{equation}

In the adiabatic limit $\omega\to0$ the relaxation and dephasing rates
(\ref{eq:T2},\ref{eq:T1}) contain the dynamical part ($\omega=0$), the geometric part
$\propto\omega$~\cite{Whitney,SMYalta}, and further non-adiabatic corrections.

We found environment-induced corrections to the phase of a two-level system beyond the adiabatic
approximation for unidirectional coupling to the environment. One can account for
more general stationary noise by adding contributions of
independent noise modes. Another case was considered in Ref.~\cite{Berger}, where the field $\B$
rotated uniformly around the $z$-axis with its horizontal component fluctuating, i.e., $\B =
B(\cos\theta\hat z + \sin\theta\;\n(t))$ and $\X=X\n(t)$ with $\n=\hat x\cos\omega
t + \hat y\sin\omega t$. In this case of `radial' noise $\n$ is stationary in the
rotating frame, and the analysis can use the same methods as above.

Apart from the short-correlated noise, one can also consider `quasi-stationary'
noise~\cite{Berger,DeChiara,Whitney}, with correlation times
longer than the time of each experimental run, $\tau_{c}\gg T$. In this case the noise $\X$ is
stationary during each run, and decoherence arises after averaging over many runs. While the
resonant part of the transverse component of $\X$ in the RF, if
present, may induce relaxation processes, to find the acquired phase and dephasing we just average
the exponential $\exp(i\int |\B'+\X| dt)$. We consider an example of uniform variation of
$\B=B_z\hat z + B_\perp(\hat x\cos\omega t + \hat y\sin\omega t)$ (and hence $\B'=\B+\omega\hat z$)
around $\hat z$ with either the `vertical' noise $\X\parallel z$ or `radial' noise~\cite{Berger}
$\X= X(\hat x\cos\omega t + \hat y\sin\omega t)$. In both cases $\X$ is stationary in the rotating
frame.

To find the average, we first expand the imaginary exponent $i\int|\B'+\X|dt$ to the second order in
$\X$: $|\B'+\X|\approx B' + X_\parallel +(X_\perp^2/2B')$. To the leading order, the phase is
obviously given by the average of the second-order term, while the dephasing by the average square
of the first-order term. Thus we find for $\n=\hat z$ the phase modification by the noise of
\begin{equation}\label{eq:phase-qs}
\delta\Phi = \frac{T}{2B'} \langle X^2\rangle \sin^2\theta'
\end{equation}
and the \emph{coherence suppression factor}
\begin{equation}\label{eq:deph-qs}
e^{-D} \quad{\rm with}\quad
D = \frac{1}{2} T^2 \langle X^2\rangle \cos^2\theta' \,.
\end{equation}
For the radial noise we find the same expressions with the sine and cosine interchanged.

From these expressions we can immediately find the limiting behavior in the near-adiabatic limit,
$\omega\to0$, of interest to Ref.~\cite{Berger}, where this limit was studied. This amounts to
expansion of the rhs of Eqs.~(\ref{eq:phase-qs},\ref{eq:deph-qs}) in $\omega$. In particular, for
the radial noise~\cite{Berger} the suppression factor is
\begin{eqnarray}
D &=&  \frac{1}{2} T^2 \langle X^2\rangle \sin^2\theta \nonumber\\
&\times& \left[
1 - \frac{2\omega}{B} \cos\theta
+ \frac{\omega^2}{B^2} (4\cos^2\theta-1)
\right] \,.
\end{eqnarray}
This reproduces the result of Ref.~\cite{Berger} except for the $\omega^2$-term, where
$4\cos^2\theta-1$ replaces $\cos^2\theta$.  This new term may be
relevant for the analysis of the differences between theory and the data in Ref.~\cite{Berger}.
To understand its origin, note that if one expands $|\B+\X|$ in the exponent only
to order $\omega$, this gives $\omega^0$, $\omega$, and $\omega^2$-terms in the suppression
factor~\cite{Berger}, however, the $\omega^2$-term in $|\B+\X|$ also contributes.

In summary, we analyzed the influence of weak dissipative environment on the Aharonov-Anandan
non-adiabatic geometric phase and dephasing. We found both the environment-induced phase
modification and dephasing for short-correlated noise and for quasi-stationary noise.

This research was supported by RSF under grant No.~14-12-00898. YM thanks M.~Cholascinski and
R.~Chhajlany for discussions at the inital stages of this work.

\end{document}